\documentstyle{article}
\begin{document}
\noindent
\begin{center}
{\large\bf ESSAY REVIEW\\[0.5cm] 
Boltzmann-Gibbs Entropy Versus Tsallis Entropy: 
Recent Contributions to Resolving the Argument of Einstein Concerning ``Neither Herr Boltzmann nor Herr Planck has given a definition of W''?}\\[0.5cm]
H.J. Haubold\\ 
Office for Outer Space Affairs, United Nations\\
P.O. Box 500 A-1400 Vienna, Austria \\[0.3cm]
A.M. Mathai\\ 
Department of Mathematics and Statistics, McGill University,\\ 
Montreal, Canada H3A 2K6 \\[0.3cm]
R.K. Saxena\\
Department of Mathematics and Statistics, Jai Narain Vyas University, Jodhpur 342001, India\\[0.3cm]
\end{center} 
{\bf Abstract.} Classical statistical mechanics of macroscopic systems in equilibrium is based on Boltzmann's principle. Tsallis has proposed a generalization of Boltzmann-Gibbs statistics. Its relation to dynamics and nonextensivity of statistical systems are matters of intensive investigation and debate. This essay review has been prepared at the occasion of awarding the 'Mexico Prize for Science and Technology 2003' to Professor Constantino Tsallis from the Brazilian Center for Research in Physics.\\[0.3cm]

Mathematical structures entered the development of physics, and problems emanating from physics influenced developments in mathematics. Examples are the role of Riemann's differential geometry in Einstein's general relativity, the dynamical theory of space and time, and the influence of Heisenberg's quantum mechanics in the development of functional analysis built on the understanding of Hilbert spaces. A prospective similar development occurred more than a decade ago when non-Abelian gauge theories emerged as the quantum field theories for describing fundamental-particle interactions (Greene, 2000). Recently, attention has turned to the applications of Riemann-Liouville fractional calculus and Mandelbrot's fractal geometry of nature to physics, including the emerging Tsallis statistical mechanics (Abe and Okamoto, 2001; Gell-Mann and Tsallis, 2004; Hilfer, 2000). The latter is the subject of this essay review.  

Statistical mechanics concerns mechanics (classical, quantum, special or general relativistic) and the theory of probabilities through the adoption of a specific entropic functional. Connection with thermodynamics and its macroscopic laws is established through this functional. The entropy concept of statistical mechanics has also been central to the creation of information theory by Shannon. A crucial step in these development was the development of a first form of non-equilibrium statistical mechanics (Prigogine, 1997).

The Maxwell-Boltzmann distribution is Gaussian in particle velocities and the whole distribution has physical meaning, so that being away from the mean is not an error. In fact the whole distribution is (i) a prediction of the Boltzmann-Gibbs entropic functional consistent with $S = k\; lnW$ and (ii) a solution of the Boltzmann equation for thermodynamic equilibrium. Tsallis entropy generalizes Boltzmann-Gibbs entropy and involves a power law: For an isolated system, $W$ is raised to the power $(1-q)$, where $q$ is the entropic parameter. But when the entropic parameter $q$ equals one, Tsallis entropy equals the logarithmic Boltzmann entropy
\begin{equation}
S_{ \mbox{(Tsallis)}} = k\; \frac{W^{1 - q} - 1}{1 - q},
\end{equation}
for $q\rightarrow 1$ one obtains
\begin{equation}
S_{ \mbox{(Boltzmann-Gibbs)}} = k\; lnW.
\end{equation}
Although the equation on Boltzmann's grave at the Vienna Central Cemetary captures his insight into entropy, he never wrote it down himself. It was Planck who, in 1900, first put it into the form that became Boltzmann's epitaph and supported the birth of quantum theory (Pais, 1982; Sch\"{o}pf, 1978). Later, in one of his papers in his miraculous year 1905, Einstein called it Boltzmann's principle (Pais, 1982; Stachel, 1998). Equation (2) reflects the fundamental insight that the second law of thermodynamics can be understood only in terms of a connection between entropy and probability and thus the second law is statistical in nature.

Einstein's perspective on classical statistical mechanics and particularly on Boltzmann's principle is reflected by his written words "Gew\"{o}hnlich wird $W$ gleichgesetzt der Anzahl der m\"{o}glichen verschiedenen Arten (Komplexionen), in welchen der ins Auge gefasste, durch die beobachtbaren Parameter eines Systems im Sinne einer Molekulartheorie unvollst\"{a}ndig definierte Zustand realisiert gedacht werden kann. Um $W$ berechnen zu k\"{o}nnen, braucht man eine {\it vollst\"{a}ndige} Theorie (etwa eine vollst\"{a}ndige molekular-mechanische Theorie) des ins Auge gefassten Systems. Deshalb erscheint es fraglich, ob bei dieser Art der Auffassung dem Boltzmannschen Prinzip {\it allein}, d.h. ohne eine {\it vollst\"{a}ndige} molekular-mechanische oder sonstige die Elementarvorg\"{a}nge vollst\"{a}ndig darstellende Theorie (Elementartheorie) irgend ein Sinn zukommt. Die Gleichung $S = (R/N)\; lgW + konst.$ erscheint ohne Beigabe einer Elementartheorie oder - wie man es auch wohl ausdr\"{u}cken kann - vom ph\"{a}nomenologischen Standpunkt aus betrachtet inhaltslos" (Einstein, 1910). ["Usually $W$ is put equal to the number of complexions in which the state of a system under consideration, uncompletely defined in the sense of a molecular theory through observable parameters, could be realized. In order to calculate $W$, one needs a {\it complete} (molecular-mechanical) theory of the system under consideration. Therefore, it is dubious whether the Boltzmann principle as such has any meaning without a {\it complete} molecular-mechanical theory or some other theory that describes the elementary processes. The equation $S = (R/N)\; lgW + const.$ seems without content, from a phenomenological point of view, without giving, in addition, such an elementary theory"].

Further, Einstein emphasizes with respect to the equation $S = (R/N)\; lgW + const.$, that "Weder Herr Boltzmann noch Herr Planck haben eine Definition von $W$ gegeben. Sie setzen rein formal $W$ = Anzahl der Komplexionen des betrachteten Zustandes" (Einstein, 1909). ["Neither Herr Boltzmann nor Herr Planck gave a defintion of $W$. They put formally $W$ = number of complexions of the state under consideration"].

The total entropy of an ensemble is some function of $W$, the number of microstates or ``complexions'',  
\begin{equation}
W = \frac{n!}{\prod_ m n_m!}
\end{equation}
and all that remains is to find that functional relationship which leads to results that agree with experiment. The relationship which leads to success identifies the ensemble entropy with $lnW$ according to Boltzmann's equation $S = k\; lnW$, where $S$ is the entropy of the entire ensemble. This equation is the fundamental principle of Boltzmann-Gibbs statistical mechanics. There is no way to prove it. One can only guess that it is right and then test the consequences against experiment. The remaining steps are mathematical. First, one uses the equation for $W$ (mindfully observing Einstein's argument on Boltzmann's and Planck's approach as given above) to get an expression for $lnW$, and then one simplifies this expression and substitutes the result into $S = k\; lnW$ to get an expression for entropy. Finally, one maximizes the entropy to obtain the preferred distribution of states. The distribution of probabilities for the possible states of an ensemble is known as the Maxwell-Boltzmann distribution. It is by no means the only distribution of probabilities that satisfies appropriate constraints on the total internal energy and total number of members of the ensemble 
\begin{equation}
U = \sum_m n_m\;\;E_m= constant,
\end{equation}
\begin{equation}
N = \sum_m n_m= constant.
\end{equation}
The constancy of $E$ results from the first law of thermodynamics. The constancy of $N$ results because the ensemble is considered to be isolated. It is the particular distribution that maximizes the ensemble entropy and which therefore conforms to the laws of thermodynamics. To be specific, the above procedure applied to Tsallis entropy in eq. (1) leads to Tsallis's distribution
\begin{equation}
f(E,q)=A_q[1-(q-1)\frac{E}{kT}]^{\frac{1}{q-1}},
\end{equation}
where
$$A_q=(q-1)^{1/2}\;\frac{3q-1}{2}\;\frac{1+q}{2}\;\frac{\Gamma(\frac{1}{2}+\frac{1}{q-1})}{\Gamma(\frac{1}{q-1})}\;A_1$$
as it leads to Maxwell-Boltzmann's distribution if applied to eq. (2)
\begin{equation}
f(E) = A_1 exp\left\{-\frac{E}{kT}\right\}
\end{equation}
where
$$A_1 = (\frac{\mu}{2\pi kT})^{3/2}$$

There is no need here to detail the record of successful applications of (7) to experiment since Boltzmann established his theory more than hundred years ago. However, it shall be highlighted here that Tsallis statistics has been applied with increasing success to an impressive range of phenomena in physics, chemistry, biology, economics, linguistics, medicine, geophysics, computer science, to name a few (Abe and Okamoto, 2001; Gell-Mann and Tsallis, 2004). Heated debates in the scientific literature confirm this and there is hope that they will produce more light for the benefit of understanding and applying Tsallis statistics \footnote{Tsallis statistics was part of the scientific programmes of the series of annual UN/ESA Workshops on Basic Space Science since their inception in 1991, particularly related to the gravitationally stabilized solar fusion reactor and solar neutrinos. A report on selected aspects of these workshops was recently given in W. Wamsteker, R. Albrecht, and H.J. Haubold (Eds.): Developing Basic Space Science World Wide: A Decade of UN/ESA Workshops, Kluwer Academic Publishers, Dordrecht 2004.}(Tsallis, 2004).\par
\bigskip
\noindent
{\bf References}\par
\bigskip
\noindent
S. Abe and Y. Okamoto (Eds.): {\it Nonextensive Statistical Mechanics and Its Applications}, Springer-Verlag, Berlin Heidelberg 2001.\\[0.3cm]
A. Einstein: Zum gegenw\"{a}rtigen Stand des Strahlungsproblems, {\it Physikalische Zeitschrift} {\bf 10} (1909) 185-193. In J. Stachel (Ed.): {\it The Collected Papers of Albert Einstein, The Swiss Years: Writings, 1900-1909}, Volume 2, Princeton University Press, Princeton 1993, pp. 541-553.\\[0.3cm]
A. Einstein: Theorie der Opaleszenz von homogenen Fl\"{u}ssigkeiten und\\ Fl\"{u}ssigkeitsgemischen in der N\"{a}he des kritischen Zustandes, {\it Annalen der Physik (Leipzig)} {\bf 33} (1910) 1275-1298. In M.J. Klein, A.J. Kox, J. Renn, and Robert Schulmann (Eds.): {\it The Collected Papers of Albert Einstein, The Swiss Years: Writings, 1909-1911}, Volume 3, Princeton University Press, Princeton 1993, pp. 286-312.\\[0.3cm]
M. Gell-Mann and C. Tsallis (Eds.): {\it Nonextensive Entropy - Interdisciplinary Applications}, Oxford University Press, New York 2004.\\[0,3cm]
B. Greene: {\it The Elegant Universe - Superstrings, Hidden Dimensions, and the Quest for the Ultimate Theory}, Vintage Books, New York 2000.\\[0.3cm]
R. Hilfer (Ed.): {\it Applications of Fractional Calculus in Physics}, World Scientific, Singapore 2000.\\[0.3cm]
A. Pais: {\it 'Subtle is the Lord ...': The Science and the Life of Albert Einstein}, Oxford University Press, Oxford 1982.\\[0.3cm]
I. Prigogine: {\it The End of Certainty - Time, Chaos, and the New Laws of Nature}, The Free Press, New York 1997.\\[0.3cm]
H.-G. Sch\"{o}pf: {\it Von Kirchhoff bis Planck} (in German), Akademie-Verlag, Berlin 1978.\\[0.3cm]
J. Stachel (Ed.): {\it Einstein's Miraculous Year - Five Papers That Changed the Face of Physics}, Princeton University Press, Princeton 1998.\\[0.3cm]
C. Tsallis: Some thoughts on theoretical physics, {\it arXiv:cond-mat/0312699, v2, 7 February 2004}.\\[1cm]
\end{document}